\documentclass[pre,preprintnumbers,amsmath,amssymb]{revtex4}
\usepackage{amssymb}
\usepackage{latexsym}
\usepackage{epsfig}

\begin{document}

\title{Thermodynamic analysis of the static spherically symmetric field equations in Rastall theory}
\author{Hooman Moradpour$^1$\footnote{h.moradpour@riaam.ac.ir} and Ines.~G. Salako$^{2,3,4}$\footnote{inessalako@gmail.com}}
\address{$^1$Research Institute for Astronomy and Astrophysics of Maragha (RIAAM), P.O. Box 55134-441, Maragha, Iran\\
$^2$African Institute for Mathematical Sciences(AIMS), $6$
Melrose Road, Muizenberg, $7945$, South Africa\\
$^3$Institut de Math\'ematiques et de Sciences Physiques
 (IMSP), Universit\'e de Porto-Novo, 01 BP 613 Porto-Novo, B\'enin \\
$^4$D\'epartement de Physique,
 Universit\'e d'Agriculture de K\'etou, BP 13 K\'etou,  B\'enin}

\begin{abstract}
The restrictions on the Rastall theory due to apply the Newtonian
limit to the theory are derived. In addition, we use the zero-zero
component of the Rastall field equations as well as the unified
first law of thermodynamics to find the Misner-Sharp mass content
confined to the event horizon of the spherically symmetric static
spacetimes in the Rastall framework. The obtained relation is
calculated for the Schwarzschild and de-Sitter back holes as two
examples. Bearing the obtained relation for the Misner-Sharp mass in
mind together with recasting the one-one component of the Rastall
field equations into the form of the first law of thermodynamics, we
obtain expressions for the horizon entropy and the work term.
Finally, we also compare the thermodynamic quantities of system,
including energy, entropy and work, with their counterparts in the
Einstein framework to have a better view about the role of the
Rastall hypothesis on the thermodynamics of system.
\end{abstract}
\maketitle

\section{Introduction\label{Intr}}
For the first time, Jacobson could use thermodynamics to derive the
Einstein field equations \cite{jacob}. The Bekenstein entropy is the
backbone of the Jackobson's approach confirming that the Einstein
theory corresponds with the Bekenstein entropy \cite{jacob}.
Generalization of his approach to $f(R)$ gravity shows that the
terms other than the Einstein tensor in the gravitational field
equations produce entropy and therefore modify the Bekenstein limit
of the horizon entropy \cite{jacob1}. In fact, such terms and thus
their corresponding entropy production terms are the signals of the
non-equilibrium thermodynamic aspects of spacetime \cite{jacob1}.

In order to study the mutual relation between gravity and
thermodynamics, we need a proper energy definition, and it seems
that the generalization of the Misner-Sharp mass in various theories
is a suitable candidate for this aim
\cite{ms,ref0,ref1,ref2,pad,pad1,pad2,pad3}. It is also shown that,
in various gravitational theory, if the gravitational field
equations are considered as the first law of thermodynamics, then we
can find an expression for the horizon entropy in the spherically
symmetric static spacetimes \cite{pad,pad1,pad2,pad3}. This approach
is used to investigate the mutual relation between the gravitational
field equations and the system thermodynamic properties, such as
entropy etc., in various gravitational theories
\cite{j1,j2,j3,j4,j5,j6,j7,j8,j9,j10,j11,j12,j13}. In all of the
above mentioned
attempts~\cite{jacob,jacob1,ms,ref0,ref1,ref2,pad,pad1,pad2,pad3,j1,j2,j3,j4,j5,j6,j7,j8,j9,j10,j11,j12,j13},
authors have only studied theories in which geometry and matter
fields are coupled to each other in a minimal way, and therefore,
the energy-momentum conservation law is met in their studies. In
fact, in their studies, the system lagrangian is equal to the sum of
the its constitutes lagrangian including the geometry and the matter
fields.

Rastall \cite{rastall} and curvature-matter coupling
\cite{od1,od2,cmc,cmc1,cmc2} theories are two gravitational theories
in which geometry and matter fields are coupled to each other in a
non-minimal way. In these theories, the lagrangian of a
gravitational system is not only a simple sum of the geometry and
the matter fields lagrangians \cite{cmc2,smal}. For these theories,
the energy-momentum conservation law is not always valid, and in
fact, the divergence of the energy-momentum tensor is proportional
with the derivative of Ricciscalar. This mutual relation between the
divergence of the energy-momentum tensor and Ricciscalar is the
backbone of the Rastall theory allowing a flux of energy between the
energy-momentum source and the geometry. The Rastall hypothesis also
modifies the Einstein field equations, a result that increases the
hopes to describe the current phase of the universe expansion
\cite{prd}. From classical point of view, the Rastall's theory may
be supported by the matter production process in the cosmos
\cite{motiv1,motiv2,motiv3,motiv4}. The mutual relation between the
Rastall cosmology and thermodynamics has recently been studied
\cite{plb}. More studies on this theory can also be found in
Refs.~\cite{neutrast,obs1,obs2,rastbr,rastsc,rascos1,rasch,more1,more2,more3,more4,cosmos3,rw1,fr,hms}.

Here, we are interested in studying the mutual relation between the
thermodynamics first law and the Rastall field equations. In fact,
we want to study the effects of the Rastall hypothesis on the
thermodynamic properties of the spherically symmetric static
spacetimes. In order to achieve this goal, generalizing the
Misner-Sharp mass of the spherically symmetric static spacetimes to
Rastall theory, we find an expression for the horizon entropy of the
spherically symmetric static spacetimes. We also compare the
thermodynamic quantities of system, including energy, entropy and
work, with their counterparts in the Einstein framework to have a
better view about the role of the Rastall hypothesis on the
thermodynamics of system. The $G=\hbar=c=1$ units are considered in
this paper.

The paper is organized as follows. In the next section, after
referring to the Rastall theory and the Newtonian limit constraints
on this theory parameters, we use the unified first law of
thermodynamics together with the zero-zero component of the Rastall
field equations to find the Misner-Sharp mass confined to en event
horizon of the static spherically symmetric spacetimes in Rastall
theory. Thereinafter, in the third section, recasting the one-one
component of the Rastall field equations into the form of the first
law of thermodynamics, we find an expression for the horizon
entropy. The comparison of the obtained thermodynamic quantities
with their counterparts in the Einstein general relativity is also
addressed in this section. The last section is devoted to concluding
remarks.

\section{Rastall field equations and the Misner-Sharp mass}
For the Rastall's original field equations, we have \cite{rastall}
\begin{eqnarray}\label{r0}
T^{\mu \nu}_{\ \ ;\mu}=\lambda R^{,\nu},
\end{eqnarray}
which finally leads to
\begin{eqnarray}\label{r1}
G_{\mu \nu}+\kappa\lambda g_{\mu \nu}R=\kappa T_{\mu \nu},
\end{eqnarray}
where $\lambda$ and $\kappa$ are the Rastall's parameter and
Rastall's gravitational coupling constant, respectively. As Rastall
has been shown \cite{rastall}, this equation leads to
$R(4\kappa\lambda-1)=T$, indicating that, since $T$ is not always
zero, the $\kappa\lambda=\frac{1}{4}$ case is not allowed in this
theory. Besides, it is also shown that if we use the Newtonian limit
and define the Rastall dimensionless parameter
$\gamma=\kappa\lambda$, then for the Rastall gravitational coupling
constant ($\kappa$) and the Rastall original parameter ($\lambda$)
we have
\begin{eqnarray}\label{gamma1}
\kappa=\frac{4\gamma-1}{6\gamma-1}8\pi,\ \
\lambda=\frac{\gamma(6\gamma-1)}{(4\gamma-1)8\pi},
\end{eqnarray}
indicating that the Einstein result ($\kappa=8\pi$) is obtainable in
the appropriate limit $\lambda=0$ which is parallel to the
$\gamma=0$ limit \cite{hms}. It is useful to mention here that as
this equation shows, since the Rastall gravitational coupling
constant diverges at $\gamma=\frac{1}{6}$, the $\gamma=\frac{1}{6}$
case is not also allowed. Finally, the Rastall's field equations can
be written as \cite{hms}
\begin{equation}\label{eins21}
G_{\mu}^\nu+\gamma g_\mu^\nu R=\frac{4\gamma-1}{6\gamma-1}8\pi
T_\mu^\nu,
\end{equation}
which leads to $R(6\gamma-1)=8\pi T$ meaning that, in agreement
with~(\ref{gamma1}), the $\gamma=\kappa\lambda=\frac{1}{6}$ case is
not allowed in this theory. Moreover, as it is obvious
from~(\ref{gamma1}), $\lambda$ diverges at $\gamma=\frac{1}{4}$, and
thus, the $\gamma=\frac{1}{4}$ case is not also allowed. Therefore,
Newtonian limit indicates that, in fact, both the
$\gamma=\frac{1}{6}$ and $\gamma=\frac{1}{4}$ cases are not allowed.
In order to study the mutual relation between the first law of
thermodynamics and the Rastall field equations, we need to
generalize the Misner-Sharp energy in this theory
\cite{ms,ref1,ref2}. It is worthwhile mentioning that one can either
use the conserved charge method or the unified first law of
thermodynamics in order to find the Misner-Sharp mass in a
gravitational theory \cite{ref1,ref2}. Here, since we are interested
in having a fully thermodynamic analysis, we use the unified first
law of thermodynamics to obtain an expression for the Misner-Sharp
mass. Consider a spherically symmetric static spacetime, with a
horizon located at $r_h$, described by
\begin{equation}\label{met2}
ds^{2}=-f(r)dt^{2}+\frac{dr^{2}}{f(r)}+r^{2}d\Omega^{2}.
\end{equation}
and filling by a source of $T^\mu_\nu$, the work density and energy
supply vector are defined as
\begin{eqnarray}\label{ce1}
W&=&-\frac{h^{ab}T_{ab}}{2}\ \ \textmd{and} \\
\Psi_a&=&T^b_a\partial_b r + W
\partial_a r, \nonumber
\end{eqnarray}
respectively. In the above equations, $d\Omega^2$ is the line
element on the two-dimensional sphere with radius $r$, and $h_{ab}$
is metric on the two-dimensional hypersurface of $(t,r)$. In order
to generalize the Misner-Sharp mass definition to the Rastall
theory, we follow the approach of \cite{ref2}, and assume that the
unified first law of thermodynamics (UFL) is valid meaning that
\begin{eqnarray}\label{ufl}
dE\equiv A\Psi_a dx^a + W dV,
\end{eqnarray}
in which $x^a$ denotes the coordinate on a two-dimensional
hypersurface of metric $h_{ab}$, and $A$ is the area of the system
boundary and therefore $A=4\pi r^2$. It is straightforward to show
that
\begin{eqnarray}\label{ufl10}
dE=\rho\ (4\pi r^2)\ dr.
\end{eqnarray}
Using the zero-zero component of Eq.~(\ref{eins21}) in rewriting
this equation, we obtain
\begin{eqnarray}\label{ufl101}
dE=\frac{6\gamma-1}{2(4\gamma-1)}[(1-2\gamma)(1-\frac{d(rf(r))}{dr})+\gamma\frac{d(r^2f^{\prime}(r))}{dr}]dr,
\end{eqnarray}
which finally leads to
\begin{eqnarray}\label{ufl101}
E=\frac{(6\gamma-1)r}{2(4\gamma-1)}[(1-2\gamma)(1-f(r))+\gamma
rf^{\prime}(r)],
\end{eqnarray}
where $\prime$ denotes derivative with respect to $r$, for the
energy confined to the radius $r$. In obtaining this result, we used
$rf^{\prime}(r)=\frac{d(rf(r))}{dr}-f(r)$. For a black hole with an
event horizon located at $r_h$, since $f(r_h)=0$, we find
\begin{eqnarray}\label{ufl2}
E=\frac{6\gamma-1}{2(4\gamma-1)}[(1-2\gamma)r_h+\gamma r_h^2
f^{\prime}(r_h)].
\end{eqnarray}
In fact, it is the Misner-Sharp mass content confined to the
mentioned horizon in the Rastall framework. The above equation can
also be written as $E=E_0(1+\Gamma)$, where
\begin{eqnarray}\label{ufl20}
E_0&=&\frac{r_h}{2},\\
\Gamma&=&\frac{\gamma}{4\gamma-1}[(6\gamma-1)r_h
f^{\prime}(r_h)+4(1-3\gamma)].\nonumber
\end{eqnarray}
It is also worthwhile mentioning that, in the $\gamma\rightarrow0$
limit, we have $\tilde{E}\equiv E_0\Gamma\rightarrow0$ leading to
$E\rightarrow E_0$ which is nothing but the Misner-Sharp mass
content of the Einstein theory \cite{pad,ms}, meaning that, as
expected, the Einstein result is recovered in the appropriate limit
of $\gamma\rightarrow0$. The $E_0\Gamma$ term comes from the Rastall
original hypothesis~(\ref{r0}) that admits a mutual energy exchange
between spacetime and energy-momentum source supporting the
geometry. A hypothesis which leads to coupling the geometry and
matter fields to each other in a non-minimal way and thus, leads to
the covariantly non-conservation of Rastall gravity. As two
examples, calculations for the value of $\Gamma$ in the
Schwarzschild and de-Sitter spacetimes lead to
\begin{eqnarray}
f(r)&=&1-\frac{2m}{r}\ \rightarrow\ r_h=2m\ \Rightarrow\ \Gamma_{sch}=\frac{3\gamma(1-2\gamma)}{4\gamma-1},\nonumber\\
f(r)&=&1-\Lambda r^2\ \rightarrow\ r_h=\frac{1}{\sqrt{\Lambda}}\
\Rightarrow\ \Gamma_{deS}=-6\gamma,
\end{eqnarray}
respectively. Finally, it is worthwhile mentioning that, in order to
have positive energy, we should have $\Gamma\geq-1$ leading to the
$r_h
f^{\prime}(r_h)\leq\frac{1+4\gamma(3\gamma-2)}{\gamma(6\gamma-1)}$
condition for the Rastall dimensionless parameter.
\section{Horizon entropy}
Here, following the approach of Ref.~\cite{pad}, we recast the
one-one component of the Rastall field equations to the form of the
first law of thermodynamics and use the result of previous section
to get the horizon entropy. The one-one component of~(\ref{eins21})
yields
\begin{equation}\label{eins24}
G_{1}^{1}+\gamma R=\frac{4\gamma-1}{6\gamma-1}8\pi T_1^1,
\end{equation}
where $T_1^1\equiv P(r)$ is the radial pressure of the
energy-momentum source \cite{pad}, and therefore, it finally takes
the
\begin{eqnarray}\label{eins25}
P(r)=\frac{6\gamma-1}{(4\gamma-1)8\pi}(\frac{1}{r^2}\left[r
f^\prime(r)
-1+f(r)\right]-\frac{\gamma}{r^2}[r^2f^{\prime\prime}(r)+4rf^{\prime}(r)-2+2f(r)]),
\end{eqnarray}
form. Here, prime ($\prime$) denotes the derivative with respect to
radius ($r$). On the event horizon, $f(r_h)=0$ and therefore,
\begin{eqnarray}\label{eins26}
P(r_h)=\frac{6\gamma-1}{(4\gamma-1)8\pi}
(\frac{1}{r_h^2}\left[r_hf^\prime(r_h) -1\right]-
\frac{\gamma}{r_h^2}[r_h^2f^{\prime\prime}(r_h)+4r_hf^\prime(r_h)-2]).
\end{eqnarray}
Multiplying this equation by $dV=4\pi r_h^2dr_h$, one gets
\begin{eqnarray}\label{eins27}
P(r_h)dV=\frac{6\gamma-1}{(4\gamma-1)}
\frac{f^\prime(r_h)}{4\pi}d(\frac{A}{4})-\frac{6\gamma-1}{2(4\gamma-1)}dr_h[1+\gamma(r_h^2f^{\prime\prime}(r_h)+4r_hf^\prime(r_h)-2)],
\end{eqnarray}
where $A=4\pi r_h^2$. For the second term of the RHS of this
equation we reach
\begin{eqnarray}\label{eins270}
\frac{6\gamma-1}{2(4\gamma-1)}[1+\gamma(r_h^2f^{\prime\prime}(r_h)+4r_hf^\prime(r_h)-2)]dr_h=\frac{6\gamma-1}{2(4\gamma-1)}[(1-2\gamma)dr_h+\gamma
d(r_h^2 f^{\prime}(r_h))+2r_h f^{\prime}(r_h)dr_h].
\end{eqnarray}
Now, bearing the $f(r_h)=0$ condition in mind, since $r_h
f^{\prime}(r_h)=(\frac{d[rf(r)]}{dr}-f(r))_{r=r_h}$, one can
simplify the RHS of the recent equation and take integral from that
to get~(\ref{ufl2}). Therefore, Eq.~(\ref{eins27}) can be written as
follow
\begin{eqnarray}\label{eins27a}
P(r_h)dV=\frac{6\gamma-1}{4\gamma-1}
\frac{f^\prime(r_h)}{4\pi}d(\frac{A}{4})-dE.
\end{eqnarray}
Since $T=f^\prime (r_h)/4\pi$ is the horizon temperature, comparing
this equation with the first law of thermodynamics ($PdV=TdS-dE$)
\cite{pad}, one gets $dS=\frac{6\gamma-1}{4\gamma-1}d(\frac{A}{4})$
which finally leads to
\begin{eqnarray}\label{eins27b}
S=(1+\frac{2\gamma}{4\gamma-1})S_0,
\end{eqnarray}
where $S_0=\frac{A}{4}$ is the Bekenstein entropy, for the horizon
entropy. It is now obvious that, in the $\gamma\rightarrow0$ limit,
we have $\tilde{S}\equiv\frac{2\gamma}{4\gamma-1}S_0\rightarrow0$,
and thus the Einstein result (the Bekenstein entropy) is recovered.
Indeed, as authors have been shown in
Refs.~\cite{jacob1,pad,pad1,pad2,pad3,j1}, terms other than the
Einstein tensor in modified gravities modify the Bekenstein limit of
the system entropy in agreement with our result ($\tilde{S}$). In
addition, since entropy is a positive quantity, the Rastall
dimensionless parameter should meet either $\gamma<\frac{1}{6}$ or
$\gamma>\frac{1}{4}$. Here, the energy-momentum conservation law is
not valid and therefore, the energy and entropy terms differ from
those of the Einstein theory \cite{pad}. In order to have a better
view about our results, we compare our results with those of the
Einstein theory \cite{pad}. Bearing the $\tilde{S}$ and $\tilde{E}$
definitions in mind, one can rewrite Eq.~(\ref{eins27a}) as
\begin{eqnarray}\label{eins27c}
P(r_h)dV=TdS_0-dE_0+(Td\tilde{S}-d\tilde{E}).
\end{eqnarray}
$TdS_0-dE_0\equiv dW_0$ is the amount of work done during apply the
hypothetical displacement $dr_h$ to the horizon in the Einstein
framework \cite{pad}. Therefore, if we decompose the work term
($P(r_h)dV$)as $P(r_h)dV=d\tilde{W}+dW_0$, then we reach
\begin{eqnarray}\label{eins27c}
d\tilde{W}\equiv P(r_h)dV-dW_0=Td\tilde{S}-d\tilde{E},
\end{eqnarray}
which denotes the additional work done in the Rastall theory in
comparison with the Esintein theory.
Finally, it is also useful to mention here that, as a desired
result, in the absence of the Rastall term ($\lambda=\gamma=0$), we
have $d\tilde{S}=d\tilde{E}=d\tilde{W}=0$ meaning that the Einstein
result is recovered \cite{pad}.

\section{Concluding remarks}
We saw that the Rastall theory of either $\gamma=\frac{1}{6}$ or
$\gamma=\frac{1}{4}$ is not allowed due to the fact that the
Newtonian limit should be satisfied by the Rastall theory.
Additionally, we used the unified first law of thermodynamics as
well as the zero-zero component of the Rastall field equations to
generalize the Misner-Sharp mass of the static spherically symmetric
spacetimes to the Rastall theory.

Moreover, bearing the obtained Misner-Sharp mass in mind, we started
from the one-one component of the Rastall field equations, and
rewrote it as the first law of thermodynamics which helped us in
finding an expression for the horizon entropy in this theory. Our
study shows that the term other than the Einstein tensor in the
gravitational field equations, the Rastall term ($\gamma g_{\mu\nu}
R$), modifies the Bekenstein limit. We also compared the obtained
thermodynamic quantities, including entropy, energy and work, with
their counterparts in the Esintein case to have a better view about
the obtained quantities. As we saw, in the $\gamma=\lambda=0$ limit,
the results of the Einstein theory are obtainable.
\section*{Conflict of Interests}
The authors declare that there is no conflict of interest regarding
the publication of this paper.
\acknowledgments{We are so grateful to the respected referee for
valuable comments and hints. The work of H. Moradpour has been
supported financially by Research Institute for Astronomy \&
Astrophysics of Maragha (RIAAM) under research project No.1/4165-5.
Ines.~G. Salako thanks  African Institute for Mathematical
Sciences(AIMS), $6$ Melrose Road, Muizenberg, $7945$, South Africa
for partial support.}

\end{document}